\documentclass[]{aa520}

\usepackage{graphicx}
\newcommand{\hi}{\mbox{H{\sc i }}}
\newcommand{\hii}{\mbox{H{\sc~ii }}}
\newcommand{\grs}{\object{GRS 1915+105} }
\newcommand{\hiig}{\object{G 45.45+0.06} }
\newcommand{\hiigb}{\object{G 45.12+0.13} }

\begin{document}
   \title{On the optical extinction and distance of \grs }

   \author{C. Chapuis\inst{1,2}
          \and
          S. Corbel\inst{1,3}}

   \offprints{C. Chapuis cchapuis@cea.fr}

   \institute{Service d'Astrophysique, DAPNIA batiment 709, l'Orme des
 merisiers, CEA Saclay, F-91191 Gif sur Yvette cedex\\
 \email{cchapuis@cea.fr}
 \and
D\'epartement de physique, batiment Buffon,\\
Universit\'e de Versailles, 45 avenue des Etats Unis\\
F-78035 Versailles cedex
\and
Universit\'e Paris VII, 2 place Jussieu, F-75251 Paris cedex 05\\}

   \date{Submitted: 27 june 2003 Received ; accepted 21 october 2003}

   \abstract{
Based on new millimeter and radio observations, we 
 reevaluate the total hydrogen column density along the line of sight to the microquasar \grs to 
$N_{H}$~=~(3.5 $\pm$ 0.3)\,$\times$\,10$^{22}$~cm$^{-2}$. Our value is consistent with the one  
 derived from X-ray measurements, namely  (3.8 $\pm$ 0.3)\,$\times$\,10$^{22}$~cm$^{-2}$
(Ebisawa et al. \cite{ebisawa}). Using
the empirical law between the  visual extinction and the total column density of hydrogen, A$_{v}$ is found to 19.6 $\pm$ 1.7~mag. This result is
$\sim 7~$mag lower than previously thought, and therefore, a reevaluation of infrared fluxes after derredening is needed. The revisited  kinematic study allows to give a lower limit to the distance of \grs, namely 6.0~kpc. Taking into account the most accurate upper limit of distance inferred from radio  data (11.2 $\pm$ 0.8 kpc; Fender et al. \cite{fender2}) as well as this lower limit, this implies a distance of 9.0 $\pm$ 3.0 kpc .

   }

   \maketitle

%


\section{\grs and its optical extinction}

\grs is a Galactic X-ray binary  in the Aquila constellation (l=45.37$^{o}$, b=-0.22$^{o}$), discovered on 15 august 1992 by the WATCH
 all-sky X-ray
monitor on board the GRANAT satellite (Castro-Tirado et al. \cite{castro1}).

 Based on follow-up observations with the VLA, \grs was shown to be the first Galactic source with apparent superluminal motion (Mirabel \& Rodr\'{\i}guez \cite{mirabel1}). 
Using the proper motion properties of relativistic radio ejecta observed with MERLIN and under the assumption of a symmetric ejection, the
more constraining upper limit of the distance was found to 11.2 $\pm$ 0.8 kpc 
(Fender et al. \cite{fender2}).

 CO absorption lines in the infrared spectrum clearly identified the donor as a  K-M III star (Greiner et al. \cite{greiner2},\cite{greiner3}). Due to the relative motion of the companion star around the center of mass of the system, these lines were modulated by Doppler effect. The period of the system (33.5 day) and its mass function (9.5 $\pm$ 3.0~$M_{\odot}$) were obtained. The mass of the compact object was measured to 14 $\pm$ 4$M_{\odot}$, which unambiguously identify the compact object as a black hole (Greiner et al. \cite{greiner3}).

During a monitoring observation of \grs by the Ryle telescope at 15 GHz, quasi-periodic oscillations were found and associated with the soft-X-ray variations on the same
time-scale (Pooley \& Fender \cite{pooley}).
Fast infrared oscillations on time-scales of less than an hour as well as radio oscillations were interpreted as small synchrotron emitting ejections of material (Fender et al. \cite{fender3}).
 Then, X-ray dips on time-scales of minutes have been interpreted as the rapid disappearance and refill of the inner accretion disk (Belloni et al. \cite{belloni1},\cite{belloni2}). Simultaneous observations revealed ejection of relativistic plasma clouds in the form of synchrotron
flares at radio and infrared wavelengths (Mirabel et al. \cite{mirabel3}).

There is currently a strong indication that the infrared emission is related to the jet, although it cannot be ruled out that a significant fraction of the infrared flux could be attributed to the disk (Fender et al. \cite{fender3}).
Thus, the infrared magnitudes must be dereddened properly in order to rely the observations to an emission process. For that purpose, the distance and the optical extinction $A_{v}$ should be measured and their limiting uncertainties carefully calculated.

The optical extinction A$_{v}$ to the system is the main parameter used to deredden the optical and infrared fluxes. Based on a measurement of the magnitude in the $R$ band, it was first evaluated to $A_{v}\sim$ 30 mag (Mirabel et al. \cite{mirabel2}). This result and the derived $A_{K}$=3.3 mag (using $\frac{A_{v}}{A_{K}}$ from Rieke \& Lebofsky \cite{rieke}) as well, are still used for dereddening (e.g. Fender et al. \cite{fender3}, Fender \& Pooley \cite{fender4}), although the authors pointed out that these results are uncertain (Fender \& Pooley \cite{fender4}).
The position of \grs is known very accurately since the discovery of the radio counterpart
 (Mirabel et al. \cite{mirabel2}). These precise coordinates allowed to perform an \hi absorption
spectrum during a radio outburst (Rodr\'{\i}guez et al. 
\cite{rodriguez}), as well as a $^{12}$CO(J=1-0) spectrum (linked to the column density of molecular hydrogen). A visual extinction of $A_{v}$=26.5 $ \pm$1$~$mag was derived. Meanwhile, another team evaluated the optical extinction to \grs to $A_{v}$=18-24 mag (Castro-Tirado, Geballe \& Lund \cite{castro2}) with the help of the X-ray hydrogen absorption column density and the pioneering work of Gorenstein (\cite{gorenstein})  relating the optical extinction to the total hydrogen column density. Moreover, we have to note that the hydrogen column density derived from Si and Fe edges (X-ray observations with Chandra) is substantially higher than 5\,$\times$\,10$^{22}$cm$^{-2}$ (leading to an extinction of 28 mag) and should imply additional local absorption (Lee et al. \cite{lee}) although the Mg and S edges lead to consistent values of $N_{H}$ with the one of Ebisawa et al. (\cite{ebisawa}). 
 An overestimate of the extinction would allow higher dereddened fluxes in K bands and flat spectrum from radio to IR (Fuchs et al. \cite{fuchs}).
Subsequently, the knowledge of the optical extinction of \grs is of prime importance for understanding the nature of the physical processes which are involved in the infrared/optical range. 
Thus, an accurate estimate of the total extinction toward \grs is needed. In order to understand the origin of these discrepancies, we conducted millimeter observations to estimate the molecular component which represent a significant contribution to the optical extinction.

We describe our observations and method in §2. Our results are reported in §3 and their implications discussed. The conclusions are then summarized in §4.


\section{Observations and method}

\subsection{Observations}

\subsubsection{IRAM}

\begin{figure*}

\resizebox{\hsize}{!}{\includegraphics{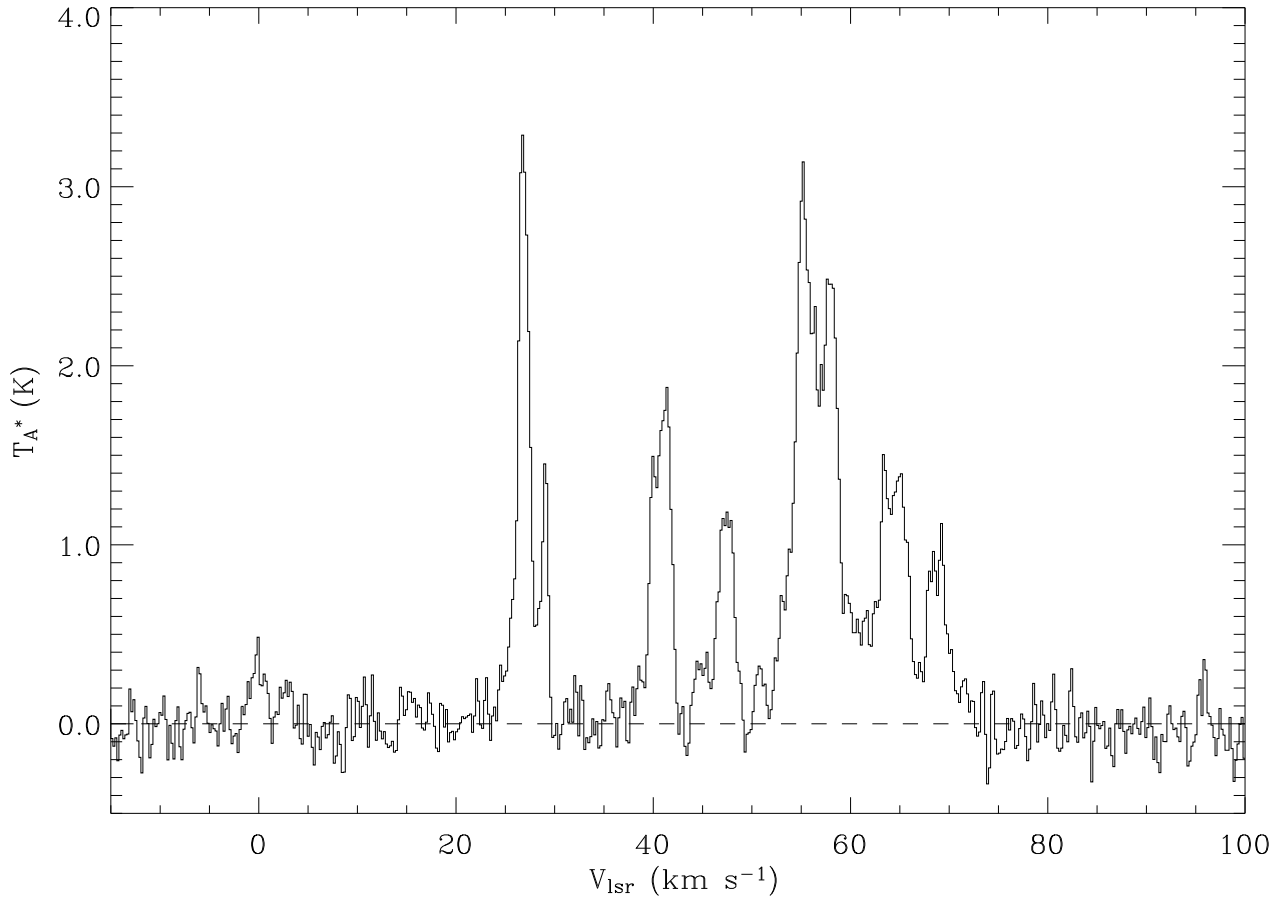}}

     \caption[ ]{ $^{12}$CO(J=1-0) antenna temperature spectrum along the line of sight of \grs performed at IRAM}
\label{COGRS}
    \end{figure*}

Using the autocorrelator as a backend connected to a 3~mm SIS receiver, we performed a high resolution $^{12}$CO(J=1-0) velocity spectrum ($\sim$~0.2~km.s$^{-1}$) at 115 GHz along the line of sight of \grs (Figure \ref{COGRS}) on 1998 june 7 at the 30 m single dish antenna of IRAM (Spain). The half power beamwidth (FWHM) of the telescope at 115.271 GHz was 27.5''. The main beam efficiency and the forward efficiency were respectively equal to 0.68 and 0.92 while a typical system temperature of $\sim$380K was observed.

   \begin{figure*}


\resizebox{\hsize}{!}{\includegraphics{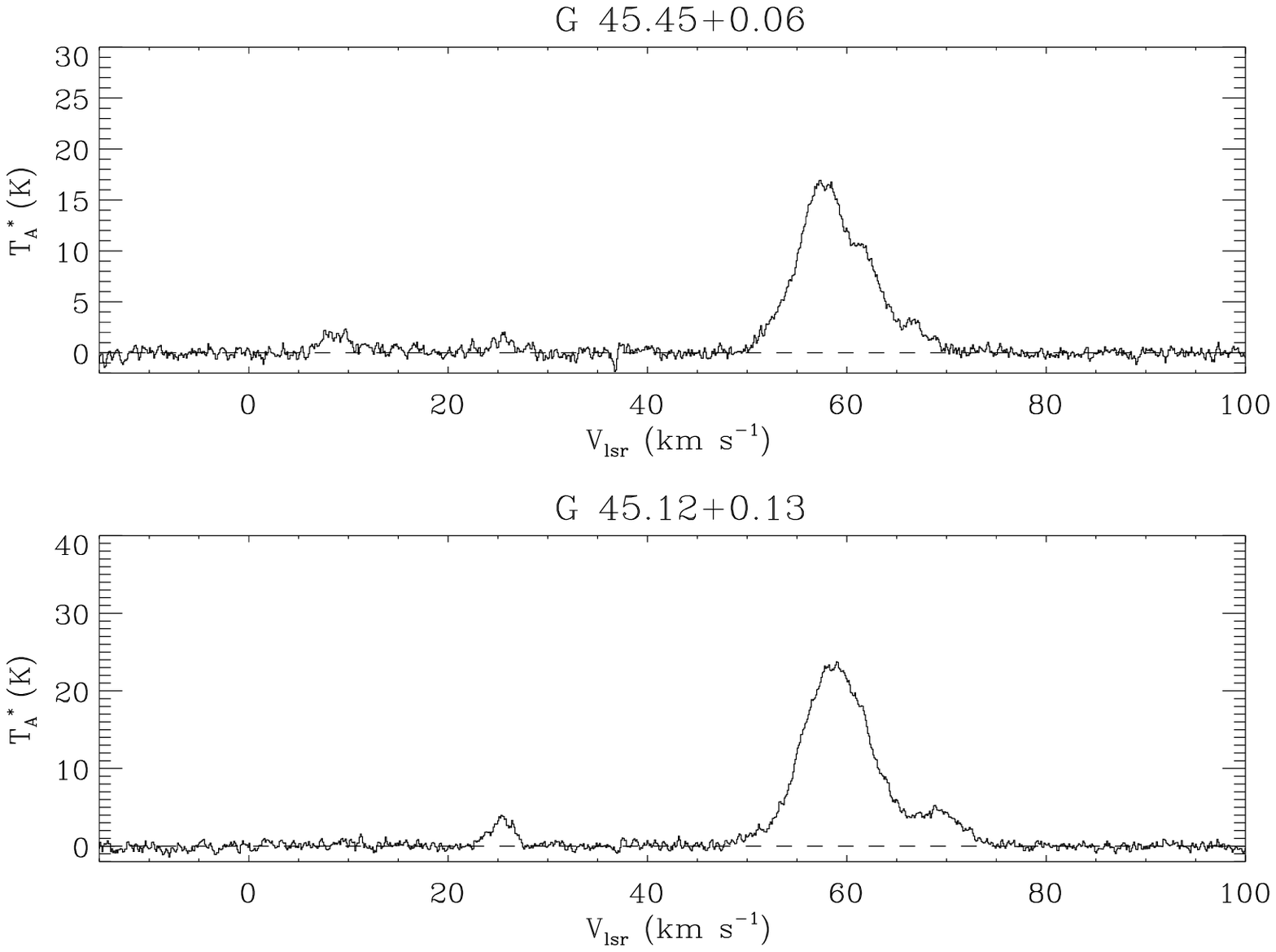}}
\hfill
     \caption[ ]{a $^{12}$CO(J=1-0) antenna temperature spectrum along the line of sight of the \hii region \hiig performed with HRS/SEST on 2 march 1999 b. same comments for \hiigb }
\label{figureCOhii}
    \end{figure*}

 The calibrated spectrum (Figure \ref{COGRS}) was fitted with gaussians to evaluate the central LSR velocity. The areas were measured by summing the antenna temperature for each channel associated to a molecular cloud. We gave names to  clouds which were simply constructed by beginning it
 with an MC (like molecular cloud) and ending it with the rough fitted LSR velocity
in km~s$^{-1}$. The characteristics of the clouds are reported in Table \ref{tableMC}.

\subsubsection{SEST}

We conducted millimeter observations with the 15 m Swedish-ESO Submillimeter Telescope (SEST) located at La Silla (Chile) on 1999 march 2. Spectra were obtained along the line of sight of \grs ($^{13}$CO(1-0) line). The FWHM beamwidth of the SEST is 45'' at 115.27 GHz and the main beam efficiency is 0.70 at this frequency. After checking that the OFF position ($\alpha(2000)=19^{h}07^{m}59.37^{s}~\delta(2000)=11^{o}08'27.1''$) was free of emission, we observed in position-switching mode. The back end was an acousto-optical spectrometer (AOS) with a frequency bandwidth of 1 GHz and a velocity resolution of 1.8 km.s$^{-1}$ at 115.27 GHz. The system was calibrated with the chopper wheel method. The system temperature during the observations was $\sim$ 650 K at 115.27 GHz for \hiig and 690 K at the same frequency for \hiigb

Spectra along the line of sight of \hii regions close to \grs were also needed to help and resolve the distance ambiguity (see next section). During the same run, we therefore performed
$^{12}$CO(J=1-0) spectra along the line of sight of two  \hii regions: \hiig and \hiigb (Figure \ref{figureCOhii}).

\subsection{Method}

 We assumed that the Galaxy rotates in a purely circular motion and
 determined the close and far distances using $R_{0}$ = 8.5 kpc and $\Theta_{0}$ = 220 km~s$^{-1}$ as standard rotation constants ($R_{0}$, galactocentric distance of the Sun and  $\Theta_{0}$, circular velocity of the Sun), using the rotation curve of Fich et al. (\cite{fich}).
 
 Our aim was to collect a maximum of information along the line of sight of
 \hii regions close to \grs as well as for \grs itself. We gathered \hi informations, CO spectra and absorption lines in order to help and locate the clouds at the most accurate distances.
 
This method was already used to evaluate the first distance to a soft gamma repeater, i.e. \object{SGR 1806-20} (Corbel et al. \cite{corbel1}, Corbel \& Eikenberry \cite{corbel3}) and repeated later after the discovery of a new soft gamma repeater, i.e. \object{SGR 1627-41} (Corbel et al. \cite{corbel2}).

 In order to convert the results from observations such as the column density of atomic hydrogen and the integrated antenna temperature of a molecular cloud in the $^{12}$CO(J=1-0) line into a total hydrogen column density and then to the optical extinction $A_{v}$, we proceed as follows.

(i) Evaluation of the integrated antenna temperature  $\int_{ }^{ } T_{A}^{*}dv$ in the 
 $^{12}$CO(J=1-0) line for each molecular cloud in the spectrum of \grs.

(ii) Converting this area into the integrated main beam temperature $W_{CO}=\int_{ }^{ } T_{mb}dv$. This conversion was done by dividing $\int_{ }^{ } T_{A}^{*}dv$ by 0.74, which is the conversion factor for IRAM. Indeed, according to the user manual of IRAM, the relation between the main beam brightness temperature $T_{mb}$ and the antenna temperature $T_{A}^{*}$ is $T_{mb}=\frac{F_{eff}}{B_{eff}}~T_{A}^{*}$, with $F_{eff}$=0.92 and $B_{eff}$=0.68  the forward efficiency and the beam efficiency respectively.

(iii) The $X_{CO}$ factor is defined as the fraction between the column density of
molecular hydrogen to the integrated area of the main beam temperature and can be written as follows: 
\begin{center}$X_{CO}=\frac{N_{H_{2}}}{W_{CO}}$
\end{center}
We adopted the value of the $X_{CO}$ factor (Strong and Mattox \cite{strong}): 
\begin{center}
$X_{CO} = \frac{N_{H_{2}}}{W_{CO}}$ = (1.9$\pm$0.2) $\times$ 10$^{20}$~K$^{-1}$~km$^{-1}$~s~cm$^{-2}$ 
\end{center}
This value was obtained by comparing $^{12}$CO(J=1-0) intensity and
gamma-ray/EGRET observations.

 Why adopting this particular value of the $X_{CO}$?
Four different methods (Dickman
\cite{dickman}, Thaddeus \& Dame \cite{thaddeus}, Bloemen et
al. \cite{bloemen}, Solomon et al. \cite{solomon2}) have led to
similar and consistent values (Solomon \& Barrett
\cite{solomon}). Moreover, the analysis of the gamma-ray data from the {\it COS-B}
observatory lead to a value of
2\,$\times$\,10$^{20}$\,K$^{-1}$~km$^{-1}$~s~cm$^{-2}$ for the $X_{CO}$ factor
(Bertsch et al. \cite{bertsch}).  The most recent data for the
Galactic diffuse gamma-ray emission are from {\it EGRET} onboard the
{\it Compton Gamma-Ray Observatory}, which is more sensitive and also
has a much higher spatial resolution. Our adopted value of $X_{CO}$ (
(1.9$\pm$0.2)\,$\times$\,10$^{20}$\,K$^{-1}$~km$^{-1}$~s~cm$^{-2}$; Strong \&
Mattox \cite{strong}) has been derived using the {\it EGRET} data.
This result is consistent with measurement at high galactic latitudes
(1.8\,$\times$\,10$^{20}$\,K$^{-1}$~km$^{-1}$~s~cm$^{-2}$; Dame et
al. \cite{dame2}). 
 Nevertheless, with the help of another
method and the same {\it EGRET} data set, a lower $X_{CO}$ -
1.6\,$\times$\,10$^{20}$\,K$^{-1}$~km$^{-1}$~s~cm$^{-2}$ - was found, which is
still consistent with a value of
$\sim$\,2\,$\times$\,10$^{20}$\,K$^{-1}$~km$^{-1}$~s~cm$^{-2}$ (Hunter et
al. \cite{hunter}).  Also, one might wonder about the influence of
metallicity, cosmic ray density and the UV radiation field on the
evaluation of the $X_{CO}$ factor.  Indeed, a decrease of the $X_{CO}$
factor could be related to a higher metallicity (Arimoto, Sofue \&
Tsujimoto \cite{arimoto}; Pilyugin \cite{pilyugin}; Boselli, Lequeux
\& Gavazzi \cite{boselli}).
Furthermore, there are evidences of
 radial variations of $X_{CO}$ (Arimoto, Sofue \& Tsujimoto
 \cite{arimoto}).
At the longitude of \grs, the
galactocentric radii of the molecular clouds on the line of sight  lies in
the range of 6 to 8.5 kpc (using the rotation curve of Fich et
 al. \cite{fich}). Therefore, if
there are radial variations of $X_{CO}$, then  the expected variation of $X_{CO}$
would be low for the line of sight to \grs.
 Moreover, in this range of galactic radii, the oxygen
 abundance ( 12 + log(O/H) = 8.9; Arimoto, Sofue \& Tsujimoto
 \cite{arimoto}), allows $X_{CO}$ = 1.95\,$\times$\,10$^{20}$\,K$^{-1}$~km$^{-1}$~s~cm$^{-2}$, according to the empirical law, log $X_{CO}$ = -1.01$\times$( 12+log(O/H)) + 29.28 ( Boselli, Lequeux
\& Gavazzi \cite{boselli}). At galactocentric distance  of 6 to 8.5 kpc, we are very far from the
central region of the Galaxy and their potential problems (Paglione et al. \cite{paglione}).
  Furthermore, our adopted
value is the mean of the interval including on one hand, the not-so
``standard'' value 2.3\,$\times$\,10$^{20}$\,K$^{-1}$~km$^{-1}$~s~cm$^{-2}$
(Strong et al. \cite{strong88}) and, on the other hand, the 
accurate value from EGRET data set to
1.6\,$\times$\,10$^{20}$\,K$^{-1}$~km$^{-1}$~s~cm$^{-2}$ (Hunter et
al. \cite{hunter}).  At last, we share Solomon \& Barrett
\cite{solomon} point of view : ``CO to H$_{2}$ conversion factor is
fairly robust - robust but not perfect''.

 Chaty et al. (\cite{chaty}) used $X_{CO}$\,
=\,3.6\,\-$\times$10$^{20}$\-\,K$^{-1}$\,km$^{-1}$\,s\,cm$^{-2}$ (Sanders, Solomon \&
Scoville \cite{sanders}) and $T_{A}^{*}$  instead of T$_{mb}$.  We calculated the
integrated antenna temperature in the $^{12}$CO(J=1-0) line for each
molecular cloud for which we considered 25 percent uncertainty of the
estimated value. These results are summarized in table \ref{tableMC}.

(iv) Inferred from a \hi absorption spectrum of \grs during an outburst, we used the total atomic hydrogen column density $N_{H{\tiny I}}=1.73$\,$\times$\,$10^{22}\frac{T_{s}}{100K}$cm$^{-2}$ (Rodr\'iguez et al. \cite{rodriguez}), where $T_{s}$ is the spin temperature.
 The total column density of hydrogen should then be evaluated, taking into account the atomic and molecular components by the calculation
 \begin{center}$N_{H}$ = $N_{H{\tiny I}}$ + 2 $N_{H_{2}}$
\end{center}

(v) Predehl \& Schmitt (\cite{predehl}) reevaluated the proportionality factor
relying empirically the visual extinction and the total column density of
hydrogen to $\frac{N_{H}}{A_{v}}$ = 1.79 10$^{21}$ cm$^{2}$ mag$^{-1}$ after the pionnering work of Gorenstein (\cite{gorenstein}). The optical extinction should then be derived from the total column density of hydrogen by the simple relation.
\begin{center}
$A_{v} = \frac{N_{H}}{ 1.79~10^{21}}$ mag
\end{center}


\section{Results}

\begin{table*}
\begin{center}\scriptsize
\begin{tabular}{ccccccccc}
\hline
 &  & & & & & & & \\
Name & LSR  & $W(CO)^{*}=$  & $N_{H_{2}}$
  & $\sigma_{N_{H_{2}}}$ & $A_{v}$ & near &  far & MC estimated\\
 &  &  $\int_{ }^{ } T_{A}^{*}dv$ &   & &  &distance & distance&distance\\
 &  &   &   & & &   &   &  \\
 & km.s$^{-1}$ &  K.km.s$^{-1}$ & cm$^{-2}$  & cm$^{-2}$ & mag & kpc & kpc & kpc\\
\hline
 &  &   &   & &  & & &\\
 \grs  &  &  &   & &  & & &\\
 &   &  &  &    &  & & &\\
 MC27 & 26.8  & 4.92  & 1.26  & 0.34 & 1.41 & 1.8 & 10.1 & 1.8\\
 &   &  &  &    &  & & &\\
 MC29 & 29.0  & 1.39  & 0.36  & 0.098 & 0.40 &  2.0 & 9.9 & 2.0\\
 &   &  &  &  &  &  & &\\
 MC41 & 40.8   & 4.25  & 1.09  & 0.30 & 1.22 &  2.9 & 9.1 &  \\
 &   &  &  &  &  &  & &\\
 MC47 & 47.4  & 2.50  & 0.64 & 0.17 & 0.72 &  3.4 & 8.6 & \\
 &   &  &  & &  &  & &\\
 MC56 & 56.4 & 13.9 & 3.57  & 0.97 & 3.99  &  4.2 & 7.7 &  \\
 &   &  &  &  &  & & &\\
 MC64 & 64.3  & 5.10  &  1.31  & 0.36 & 1.46 & 5.4 & 6.6 &  \\
 &   &  &  &  &  & & &\\
 MC69 & 68.9  & 2.20  & 0.56  & 0.15 & 0.63 & 6.0 & 6.0 & 6.0\\
 &   &  &  &  &  &  & &\\
 \hline
 &   &    &   &  &  &  & &\\
total &    & 34.3 &  8.79 & 1.4 & 9.83 &  & &\\
 &   &   &   &  & &   & &\\
\hline

\end{tabular}
\end{center}
\caption{ Summary of molecular clouds observations from $^{12}$CO(J=1-0) 
spectra with IRAM and SEST for \grs line of 
sight, where the columns represent
(1) name, (2) LSR fitted velocity in km s$^{-1}$ (3) area of the antenna temperature $T_{A}^{*}$ in
K km s$^{-1}$ (4) $N_{H_{2}}$ in 10$^{21}$ cm$^{-2}$ (5) $\sigma _{N_{H_{2}}}$ in 10$^{21}$ cm$^{-2}$
(6) optical extinction $A_{v}$ in magnitude (7)(8)(9) columns meaning is explicit}
\label{tableMC}
\end{table*}

\subsection{Extinction along the line of sight of \grs}

Our first goal is to calculate the maximum value of the total column density of hydrogen. For that purpose, we need to use all the available atomic and molecular matter along the line of sight of \grs.
Following the procedure described in the previous section, the results of our observations, i.e. velocity, distance, integrated antenna temperature and $N_{H_{2}}$ of each cloud, are summarized in table \ref{tableMC}. 
We calculated the total column density of molecular hydrogen
\begin{center}
$N_{H_{2}}$ = (8.8$\pm$1.4)\,$\times$\,10$^{21}$ cm$^{-2}$ 
\end{center}
The value of Chaty et al. (\cite{chaty}) would be consistent with this one, using the same $X_{CO}$.

The total column density of hydrogen becomes
\begin{center} 
 $N_{H}$ = (3.5$\pm$0.3)\,$\times$\,10$^{22}$ cm$^{-2}$
\end{center}
taking into account the \hi column density value of (Rodriguez et al. \cite{rodriguez}).
This value is consistent with the one of Ebisawa et al. (\cite{ebisawa})
 derived from ASCA X-ray 
observations of \grs, which is the 
most accurate to date: $N_{H}$ = (3.8$\pm$0.3)\,$\times$\,10$^{22}$ cm$^{-2}$. 
Subsequently, it is not necessary to locate each cloud accurately, since all the absorbing matter lies between the system and us.
Any possible additional absorption seen in the X-ray spectra should then have a local origin.
According to the proportionality factor
relying empirically the visual extinction and the total column density of
hydrogen (Predehl \& Schmitt \cite{predehl}), we finally found
\begin{center}
 A$_{v}$ = 19.6$\pm$1.7 mag.
\end{center}
This estimate is $\sim$~7~mag lower than 26.5 (Chaty et al. \cite{chaty}), which was based on an overevaluated $X_{CO}$ by a factor of $\sim$2. The visual extinction is still consistent with the high observed reddening of the source in the infrared and consistent with the suggested value of $\sim$20 mag (Mahoney et al. \cite{mahoney}) and with the interval 18-24 calculated earlier (Castro-Tirado, Geballe \& Lund \cite{castro2}). Our value implies that the previous dereddeneding of infrared data processed with  A$_{v}$ from 26.5 to 30 are uncorrect.

We note that a possible negligible excess around 2 km.s$^{-1}$ can be seen on Figure \ref{COGRS}, which effects are negligible on $N_{H_{2}}$ and therefore on $A_{v}$. Its velocity, confirmed by a \hi difference spectrum (Dhawan et al. \cite{dhawan}), would suggest either a solar component or a close to \grs component.
\subsection{Is it possible to constrain the distance to \grs?}

Using the rotation curve of the Galaxy (Fich et al. \cite{fich}) and the observed $^{12}$CO(J=1-0) luminosities, we estimate the far and near kinematic distances of each molecular cloud. These values are reported into the table \ref{tableMC} for the line of sight of \grs (Figure \ref{COGRS}). Since all the available atomic and molecular matter is taken into account and is consistent with the value derived from X-ray observations (i.e., $N_{H}$=(3.8$\pm$ 0.3)\,$\times$\,10$^{22}$~cm$^{-2}$,
 Ebisawa et al. \cite{ebisawa}), it is not necessary to locate the molecular cloud accurately along the line of sight in order to discuss whether \grs is closer or beyond each of them. We just have to identify the farthest cloud and to evaluate its kinematic distance accurately. This would then be the lower limit of the distance to \grs.

\hiig and \hiigb are two \hii regions, respectively 18 arcmin and 26 arcmin away from \grs. The spectrum of \hiig reveals molecular clouds around 10, 25 km.s$^{-1}$ and a complex between 50 to 70  km.s$^{-1}$ as well (Figure \ref{figureCOhii}) - FWHM$\sim$10 km s$^{-1}$-. This large complex - [46,49] - was still reported as associated to four \hii regions (Dame et al. \cite{dame}). Among them, the two \hii regions close to \grs are noted in the map Fig.5 of Dame et al. (\cite{dame}). The FWHM distribution spreading for the largest galactic molecular clouds lies typically in the range 4-15 km s$^{-1}$ and peaks around 10 km s $^{-1}$. Along the line of sight of \grs, MC27 is at the near distance (Dame et al. \cite{dame}), cf. table \ref{tableMC}. Surprisingly, no molecular matter is seen close to the 41 km.s$^{-1}$ velocity in both spectra of \hiig and \hiigb (Figure \ref{figureCOhii}) although the line of sight to \grs exhibits additionnal \hi absorption at 41$\pm$ 6 km.s$^{-1}$ relative to the line of sight to \hiig (Rodriguez et al. \cite{rodriguez}). It was therefore claimed that the additionnal \hi absorption was farther than \hiig (Rodriguez et al. \cite{rodriguez}), so that \grs was beyond \hiig and then behind a cloud located at the kinematic distance of 9.4 $\pm$ 0.5 kpc, which we consistently located at 9.1 kpc (cf. table \ref{tableMC}). This is no longer an evidence if we assume, as Rodr\'{\i}guez et al. (\cite{rodriguez}) did, that the CO feature and the \hi additional absorption onto \grs come from the same cloud, i.e. MC41. We suggest that the complex along the line of sight of \grs , for which velocities are seen from 50 to 70  km.s$^{-1}$, is an extension of the one seen along both lines of sight of the two \hii regions (Figure \ref{figureCOhii}). Thus, its location should be the tangent point. It has a large contribution to the extinction to \grs ($\sim$~6 mag), so that it could not be possible to explain the $A_{v}$ without its contribution. This is confirmed by \hi observations which allowed to put a lower distance to \grs beyond the tangent point (Rodriguez et al. \cite{rodriguez}). \grs is then behind this complex and therefore, the closest lower limit is the distance to the tangent point, i.e., the distance to this complex, namely 6.0 kpc.

Due to their non negligible contributions to the extinction (cf table \ref{tableMC}), MC41 and MC47 have to be taken into account in order to describe the extinction properly. We point out that the optical extinction should be 2~mag lower (i.e. 17.5~mag) if MC41 and MC47 are behind \grs. The implied hydrogen column density ($3.1$\,$\times$\,$10^{22}$cm$^{-2}$) is still consistent with the most accurate X-ray measurement in a two sigma confidence level (Ebisawa et al. \cite{ebisawa}), although it is at the far end of the interval and that it could mean that we need at least the contribution of one or both MC41 and MC47 to be consistent with measurements with a higher confidence level.

During 1997 October/November, multiple relativistic ejections from \grs were observed in radio 
range with MERLIN and an upper limit
of the distance to \grs was deduced to 11.2$\pm$0.8 kpc (Fender et al. 
\cite{fender2}),
which is about 1 kpc lower than the upper pionnering value of Mirabel \&  Rodr\'{\i}guez 
(\cite{mirabel1}) inferred from a similar discussion. This upper limit is consistent with \hi emission spectra in this direction of the sky (Radhakrishnan et al. \cite{radha}) because of a lack of absorption at negative LSR radial velocities (Rodriguez et al. \cite{rodriguez}, Dhawan, Goss \& Rodriguez \cite{dhawan}).

Thus, the distance of \grs
should be greater than 6.0 and less than 12.0 kpc, using a one sigma allowed interval on the upper limit of Fender et al. (\cite{fender2}). We then put \grs at 9.0$\pm$3.0 kpc. Note carefully that the 3 kpc interval is not a one sigma confidence level. We consider this expression as an interval and not as an evaluation of the distance at a one sigma significance. The
 consequence of that result, enforced by the lower $A_{v}$ we found, is that \grs is less luminous than expected in the optical and infrared ranges.

\subsection{A metal rich local dust absorption?}

Assuming solar abundances, Chandra spectral observation leaded to an H column density derived from silicon and iron K-edges ((8.4$^{+0.1}_{-0.2}$)\,$\times$\,$10^{22}$cm$^{-2}$ and (9.3$^{+1.6}_{-1.3}$)\,$\times$\,$10^{22}$cm$^{-2}$ respectively)  substantially higher than the values derived from sulphur and magnesium K-edges ((3.2$^{+0.1}_{-0.6}$)\,$\times$\,$10^{22}$cm$^{-2}$ and (3.1$^{+0.3}_{-0.3}$)\,$\times$\,$10^{22}$cm$^{-2}$ repectively)(Lee et al. \cite{lee}). The last two values are both consistent with our calculated $N_{H}$ as well as the one from ASCA (Ebisawa et al. \cite{ebisawa}). Based on the values of $N_{H}$ inferred from Si and Fe edges, it was suggested that either a highly unusual supernova or external supernova activity local to the binary could explain the overabundance in iron and silicon (Lee et al. \cite{lee}). Indeed, it seems that a 10$^{52}$erg kinetic energy SNe with high mixing, helped by a small mass cut and/or asphericity in explosion can provide high amounts of ejected $^{56}Ni$, and thus ejected iron after $^{56}Ni$ decay (Nomoto et al. \cite{nomoto}). Could this metal rich component be a relic of the SNR now diluted in the ISM? 

 We would like to note several alternative explanations since such values of the $N_{H}$ are inconsistent with our measurement of the total content along the line of sight. We propose that a Si/Fe enriched component should be either along the line of sight or local to \grs. The matter could come from a bright star (ISOGAL-PJ191511.2+105622) seen in the ISOGAL survey (Felli et al. \cite{felli}), 18'' away to the line of sight of \grs and classified as an AGB star (Fuchs et al. \cite{fuchs}) after a dereddening with $A_{v}$=20 mag.
Although the photosphere is thought to be originally solar in composition, some extremely metal poor post AGB stars (A to F stars) are depleted in elements such as Fe, Ca, Si, Cr (two orders of magnitude below solar) and with solar photospheric abundances of C, N, O, S and Zn (Mathis \& Lamers \cite{mathis}). The missing elements should be incorporated into grains that were separated from atmosphere. If this AGB is at the same distance than \grs, only little material can account to the anomalous abundances of Si/Fe (Fuchs et al. \cite{fuchs}). On the other hand, if this AGB star is much closer to the Earth than \grs, its surrounding nebula could likely have contaminate the line of sight of \grs and enriched it in heavy elements. Using the Draine extinction law (\cite{draine}) and the same $A_{v}$=20 mag as for \grs, the ISOGAL-PJ191511.2+105622 spectrum exhibits a bump at 9.7$\mu$m. Therefore, the extinction to ISOGAL-PJ191511.2+105622 would be overestimated, and then the distance closer. In order to give some solidity to this hypothesis, using the rough value of 10 mag, the bump disappears in the spectrum which becomes more likely the one of a star in this range of energy than the previous one (Fuchs, private communication). The derived dereddened  7$\mu$m and 15 $\mu$m magnitudes are respectively equal to [7]= 7.60 mag and [15]= 7.04 mag. The [15]- [7]-[15] diagram (Felli et al. \cite{felli}, Figure 1.) reenforces the AGB association. Although the value of $\sim$10 mag is a rough estimate of its optical extinction, it implies however that this AGB lies in front of the big complex located at the tangent point, and therefore with an upper limit of its distance of 6.0 kpc. The angular separation between \grs and the line of sight of this AGB is 18''. At a distance of 6 kpc, this is converted to a projected distance on the plane of the sky of 0.5 pc. Moreover, it is known that an AGB star can eject material on distances as far as 1 pc.
We thus claim that the surrounding dust of this AGB type star, closer to the Earth and with a lower extinction than \grs, could contaminate significantly the line of sight of \grs and contribute to the anomalous metallicity observed in the X-rays with Chandra (Lee et al. \cite{lee}) which cannot be taken into account with our millimiter and radio determinations of $N_{H}$ and $A_{v}$.

\begin{table*}[ht!]
\begin{center}\scriptsize
\begin{tabular}{cccccccccc}
\hline
 & & & &  &  & & & &\\
Band & U & B & V & R & I & J & H & K & L\\
\hline
 & & & & &  & & & &\\
$\lambda$ ($\mu m$) & 0.365 & 0.44 & 0.55 & 0.70 & 0.90 & 1.25 & 1.65 & 2.2 & 3.6\\
 & & & & &  & & & \\
$A_{\lambda}$ (mag) & 30.8$\pm$2.8 & 26.2$\pm$2.3 & {\bf 19.6}$\pm${\bf 1.7} & 14.7$\pm$1.3 & 9.4$\pm$0.8 & 5.5$\pm$0.5 & 3.7$\pm$0.4 & 2.2$\pm$0.3 & 1.1$\pm$0.1\\
 & & & & &  & & & &\\
\hline

\end{tabular}
\end{center}
\caption{ Summary of \grs extinctions from U to L bands. Lines
(1) Band, (2) wavelength $\lambda$ in $\mu$m (3) Value and uncertainty of $A_{\lambda}$ for each band in magnitudes }
\label{tableAv}
\end{table*}

\section{Conclusions}

The column density of hydrogen along the line of sight of \grs is found to $N_{H}$ = (3.5$\pm$0.3)\,$\times$\,10$^{22}$ cm$^{-2}$. This value is consistent with the X-ray observations of ASCA  (Ebisawa et al. \cite{ebisawa})
constraining the column density of hydrogen to ($3.8\pm 0.3$)\,$\times$\,10$^{22}$ cm$^{-2}$ .

 The visual extinction inferred from molecular spectra and \hi absorption spectrum  is 
 $A_{v}$ = 19.6$\pm$1.7~mag, 
 consistent with the guess of Castro-Tirado, Geballe \& Lund (\cite{castro2}) who found the likely range of 18-24 mag. Our value of the optical extinction is then still consistent with a high reddening of the source. Based on Cardelli et al. (\cite{cardelli}) relations between  $A_{v}$ and $A_{\lambda}$, we deduce $A_{\lambda}$ from U to L bands (Table \ref{tableAv}). In order to evaluate the uncertainty on $A_{\lambda}$, we took into account the uncertainty on $A_{v}$  as well as the spreading of the values of Cardelli et al. (\cite{cardelli}) as  uncertainty on the proportionnality factor between $A_{v}$ and $A_{\lambda}$.
The consequence is that all dereddening using extinctions of $A_{K}$=3.0 mag or more (Chaty et al. \cite{chaty}, Fender \& Pooley \cite{fender4}) should be reprocessed and their results reevaluated.

With the help of the rotation galactic rotation description (Fich et al. \cite{fich}),
 the properties of close to \grs \hii regions, $^{12}$CO(J=1-0) spectra along the
 line of sight of \grs and two close \hii regions as well as the most recent upper limit
 of distance (Fender et al. \cite{fender2}), we reevaluated the distance
 to \grs at 9.0$\pm$3.0 kpc. It is important to notice that no molecular cloud is seen at a velocity close to 40 km.s$^{-1}$ in both line of sights of the two close to \grs \hii regions, i.e. \hiig and \hiigb.

\begin{acknowledgements}

We acknowledge  Y. Fuchs and S. Chaty  for valuable discussion. We warmfully thank S. Shore for his relevant comments on the paper. We are grateful to S. Leon for conducting the IRAM observations.
This research has made use of the SIMBAD database, operated at CDS, Strasbourg, France and of NASA's Astrophysics Data System Abstract Service.

\end{acknowledgements}

\end{document}